\documentclass[aps,noshowpacs,nofootinbib]{revtex4}

\usepackage[utf8]{inputenc}
\usepackage{color}
\usepackage{amsmath,amssymb}
\usepackage{amsfonts}
\usepackage{verbatim}
\usepackage{makeidx}
\usepackage{hyperref}
\usepackage{slashed}
\usepackage[vcentermath]{youngtab}
\usepackage{float}
\usepackage{booktabs}
\usepackage{graphicx}
\usepackage{mathrsfs}
\usepackage{bm}
\usepackage{braket}
\usepackage{url}
\usepackage{etoolbox}
\apptocmd{\sloppy}{\hbadness 10000\relax}{}{}
\allowdisplaybreaks[3]
 \newcommand{\be}{\begin{equation}}
   \newcommand{\ee}{\end{equation}}
     \newcommand{\bea}{\begin{eqnarray}}
   \newcommand{\eea}{\end{eqnarray}}

\usepackage[normalem]{ulem}  
\usepackage{color}
\renewcommand\sout{\bgroup \color{red} \ULdepth=-.5ex \ULset}
\begin{document}

\title{$J/\psi$-pair resonances by LHCb: a new revolution?} 
\author{Luciano Maiani}
\address{Dipartimento di Fisica and INFN,  Sapienza  Universit\`a di Roma, Piazzale Aldo Moro 2, I-00185 Roma, Italy}
\address{T. D. Lee Institute, Shanghai Jiao Tong University, Shanghai, 200240, China}
%\date{\today}

  \maketitle

\samepage

{\it Contribution to Science Bulletin as a NEWS \& VIEWS Paper}

\begin{figure}[htb]
   \centering
   \includegraphics[width=0.6 \linewidth]{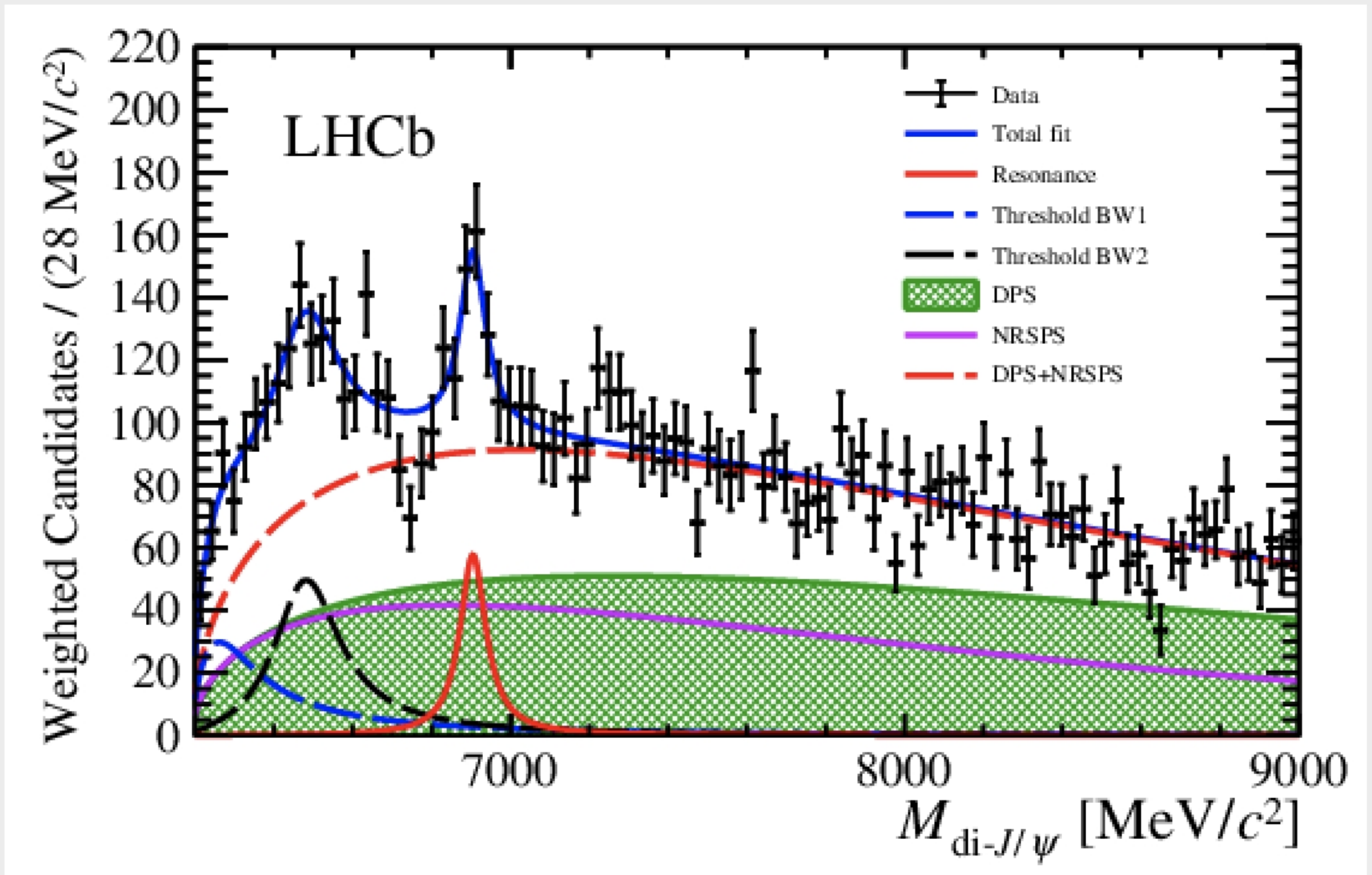}
   \caption{\footnotesize{The mass spectrum of $J/\psi$ pairs superposed by fit result of Model I, LHCb~\cite{Aaij:2020fnh}.}}
         \label{funo}
\end{figure}

\vskip01cm
In a research article appeared in this journal, the LHCb Collaboration has presented an extraordinary result on the mass spectrum of $J/\psi$-pairs produced in proton-proton high energy collisions. The spectrum shows: {\it a narrow structure around $6.9$ GeV matching the lineshape of a resonance and a broad structure just above twice the $J/\psi$ mass}~\cite{Aaij:2020fnh}. In the same region, more resonances may be present, not clearly distinguished within present statistics, (see Fig.~\ref{funo}). 

A new spectroscopy is waiting for us, that of fully charm tetraquarks, which parallels the spectroscopy of charmonia revealed by the $J/\Psi$ discovery of 1974. 
%@@@@@@@@@@@@@@@@@@@@@@@
%@@@@@@@@@@@@@@@@@@@@@@@@@@

The LHCb discovery did not come unexpected. 

LHCb has observed analogous peaks in the mass distribution of $J/\psi -\phi$ pairs from $B^+$ decay~\cite{Aaij:2016iza}. 
Theoretically, mixed $c s$ tetraquarks have been studied with the quark model~\cite{Maiani:2016wlq}
~and QCD sum rules~\cite{Chen:2017dpy}.
~The possible existence of fully heavy tetraquarks was proposed as early as 1985~\cite{Heller:1985cb}
~and it has been widely considered~\cite{Wu:2016vtq,Bedolla:2019zwg}, together with doubly heavy tetraquarks, after the observation of doubly heavy baryons.

\emph{\bf Exotic Hadrons.}~The hypothetical existence of hadronic states with more than minimal quark content 
was considered in the seminal paper, where Murray Gell-Mann, in 1964, proposed the existence of quark constituents in the hadrons: {\it Baryons can now be constructed from quarks by using the combinations ($qqq$), ($qqqq\bar q$), etc., while mesons are made out of ($q\bar q$), ($qq\bar q\bar q$), etc. \dots the lowest baryon configuration ($qqq$) gives just the representations ${\bf 1}, {\bf 8}$, and ${\bf 10}$ that have been observed, while the lowest meson configuration ($q\bar q$) similarly gives just the ${\bf 1}$ and ${\bf 8}$}  \cite{GellMann:1964nj}. 
In the same year, quarks were introduced, independently,  by George Zweig~\cite{Zweig:1964}.

In the sixties and seventies, the existence of $qq\bar q\bar q$ tetraquarks was considered unlikely due to the non-observation of $Q=+2$ mesons, that would arise from configurations such as $uu\bar d\bar d$.

With the advent of QCD, Robert Jaffe proposed a way to overcome the problem~\cite{Jaffe:1976ig}: a diquark with spin $S=0$ and colour ${\bar {\bf 3}}$ which, by Fermi statistics, has to be antisymmetric in flavour, is in a ${\bar {\bf 3}}$ of $SU(3)_{flavour}$. Binding  such {\it good} diquark \footnote{the term {\it good} diquark was coined in ~\cite{Jaffe:2004ph}, to contrast the  {\it bad} diquark, spin $S=1$ and flavour symmetric, which would give rise to really exotic mesons} with an equally good antidiquark gives rise only to non-exotic octet and singlet tetraquark mesons ({\it crypto-exotic} hadrons). 
Jaffe applied this idea to the lightest scalar mesons, described as  good diquark anti-diquark pairs,
~thereby explaining why the scalar singlet state, $\sigma=f_0(500)=[ud][\bar u\bar d]$, is lighter than its $I=0,1$ partners, $a_0(980),~f_0(980)=[sq][\bar s\bar q^\prime]$,~($q,q^\prime=u,d$) and why $f_0(980)$ has an appreciable decay into $K\bar K$, in spite of the small  phase space. 

Heavy-light good and bad diquarks have been introduced in the tetraquark descripton of $X(3872),~Z_c(3900)$ and $Z_c(4020)$, following the idea that, in the
limit of infinite charm mass, bad diquarks should also form bound states if good diquarks do so~\cite{Maiani:2004vq}.

Interestingly, the very same problem that motivated Jaffe's tetraquark solution of the properties of $f_0(980)$,~{\it i.e.} its affinity to $K$ mesons, is at the root of the alternative picture developed somewhat later, the idea that $f_0(980)$ is a $K\bar K$  bound state held together by pion exchange forces~\cite{Weinstein:1990gu}. The same valence quarks present in Jaffe's tetraquark, are rearranged into two color singlet hadrons bound by the same force that causes nuclear binding. It is a picture that had been introduced earlier by De Rujula, Georgi and Glashow~\cite{DeRujula:1976zlg} under the pictorial name of {\it hadron molecule}.

Recent years have seen the observation of  several four valence quark  states that cannot be included in the systematics of $ q\bar q$ mesons, like $Z(4430)$~\cite{Choi:2007wga,Aaij:2014jqa}, $Z_c(3900)$ and $Z_c(4020)$~\cite{Ablikim:2013wzq}. Similar particles have also been found in the bottom sector, $Z_b(10610)$ and $Z_b(10650)$, observed by the Belle collaboration \cite{Bondar}  (see~\cite{Ali:2019roi} 
~for recent reviews). In 2015, LHCb has observed resonant lines in the $J/\psi-p$ spectrum~\cite{Aaij:2015tga}, resolved more recently into three fully fledged pentaquarks~\cite{Aaij:2019vzc}.

In spite of the large body of experimental information accumulated on exotic hadrons, no consensus has been reached yet about the way  valence quarks are organised inside them.
 %@@@@@@@@@@@@@@@@@@@@@@@
 \begin{figure}[htb]
   \centering
   \includegraphics[width=0.4 \linewidth]{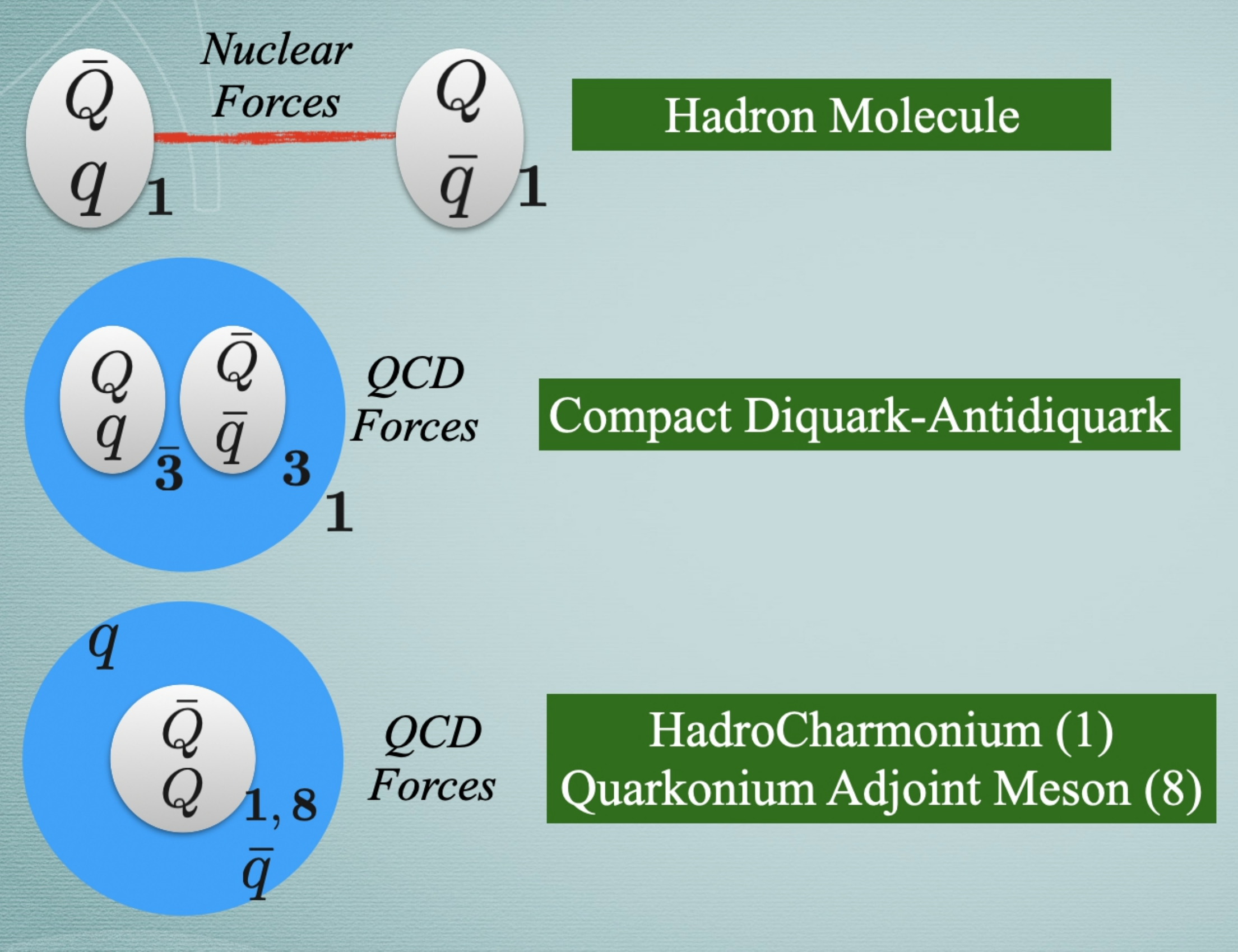}
   \caption{\footnotesize{Exotic hadrons: artist's view of the quark configurations presently under discussion. The numbers appended as subscripts correspond to the dimension of colour representations.}}
         \label{fdue}
\end{figure}
%@@@@@@@@@@@@@@@@@@@@@@@@@@
 Loosely bound molecules~\cite{Guo:2017jvc} and compact tetraquarks~\cite{Ali:2019roi} are  the two opposite extrema of a spectrum of more complex solutions to the problem. Another interesting suggestion is that of hadrocharmonia, {\it i.e.} relatively  compact charmonium embedded in a light quark mesonic excitation~\cite{Li:2013ssa},~(see  Fig.~\ref{fdue}). To complete the picture, we may add that some authors consider the possibility that exotic hadrons might simply be a threshold kinematical effect, a cusp, as detailed in \cite{Swanson:2006st}. 

\emph{\bf The ${\bf 6.9}$~GeV resonance as a Tetraquark.} There are no known color singlet, light particles that may be exchanged between two charmonia to produce binding or final state interactions. That the  ${6.9}$~GeV resonance is a hadron molecule or a threshold cusp seem to me a rather unlikely possibility. What can we say assuming that the resonance(s) seen in Fig.~\ref{funo} are  fully charm tetraquarks?

Assuming the diquark $[cc]$ in colour ${\bar{\bf 3}}$, the total spin of each diquark is $S=1$: colour antisymmetry and Fermi statistics imply a {\it bad} diquark. 

$S$-wave states have positive parity and are classified according to:
\begin{enumerate}
\item C=+1: $J^{PC}=0^{++}, 2^{++}$. $S$-wave decays: $J/\psi$-pair; $\eta_c$-pair, $\chi_{c0}(1P)$-pair ($J=0$ only), $\chi_{c1,2}(1P)$-pair ($J=0,~2$ if allowed by phase space); $D \bar D$ + light mesons;
\item C= -1: $J^{PC}=1^{+-}$. $S$-wave decays: $J/\psi+\eta_c$, $D \bar D$ + light mesons.
\end{enumerate}
In ref.~\cite{Becchi:2020uvq}, we estimate branching fractions and decay rates and give upper  bounds to the product $\sigma(pp\to {\cal T})\times BR({\cal T}\to 4\mu)$ in $pp$ collisions, where ${\cal T}$ denotes an $S$-wave, fully charm tetraquark. The branching  ratio in 4 muons is more favourable for the $2^{++}$  w.r.t. $0^{++}$, by a factor of $6$. In addition, $2^{++}$ is produced with a statistical factor 2$J$+1=5. In total
one obtains a {\it  visibility ratio} $ 2^{++} : 0^{++}=30:1$! 

Tetraquark masses have been computed by Bedolla {\it et al.}~\cite{Bedolla:2019zwg}
 treating the tetraquark as a two body, diquark-antidiquark system in QCD inspired, Coulomb plus linear, potential. Authors include computation of the energy levels of radial and orbital excitations.

The prediction includes one {\it a priori} unknown additive constant (to fix the zero of energy for confined states) which has to be determined from one mass of the spectrum. In~\cite{Bedolla:2019zwg} the constant was taken (provisionally) from calculations of meson masses. To compare with the experimental spectrum, one has to identify the state which corresponds to the ${6.9}$~GeV resonance and shift all other masses accordingly.  Needless to say, determining the spin-parity of the resonances is the first priority.

\emph{\bf Concluding.} The observation of the $6.9$ GeV peak opens several exciting possibilities and it may be a real {\it game changer}. Theoretically, there are little doubts that four heavy quarks may form bound states, at least in the infinite mass limit. The fully charm tetraquark spectrum may be amenable to theoretical calculation along the lines developed for quarkonium spectrum and the results may shed light to the sector of heavy-light tetraquarks, in the same way that quarkonium spectroscopy helped a better understanding of the light hadron spectrum (as reflected e.g. in the work of Ref.~\cite{Godfrey:1985xj}).

The $6.9$ GeV peak seems to go well with a $2^{++}$ resonance, for production rate and total width, but nothing can be said until a direct spin-parity determination is produced. More states are expected, below and above $6.9$ GeV, if the latter is indeed the $2^{++}$ $S$-wave state.
  
Another possibility is that $S$-wave states are below threshold and we are seeing the spectrum of radial and orbital excitations. Again spin-parity determinations will be crucial. 

One could also consider the  4 charm state with the diquark in color ${\bf 6}$. Quarks  do repel each other, in this case, but the overall confinement forces could keep the system bound. Individual diquark spins are $0$ and only a $J^{PC}=0^{++}$ state is expected in S-wave.

\vskip0.3cm

{\it \dots so much accomplished, and so much more left to do} (Winston Churchill).

\acknowledgements
I would like to thank Antonello Polosa and Elena Santopinto for many discussions and suggestions on exotic hadrons and for support in writing this comment. A conversation with Sheldon Stone was very useful in the preparation of the work quoted in~\cite{Becchi:2020uvq}.


\newpage

\begin{thebibliography}{}


\bibitem{Aaij:2020fnh}
R.~Aaij \textit{et al.} [LHCb], [arXiv:2006.16957 [hep-ex]].

\bibitem{Aaij:2016iza}
R.~Aaij \textit{et al.} [LHCb], Phys. Rev. Lett. \textbf{118} (2017), 022003. 

\bibitem{Maiani:2016wlq}
L.~Maiani, A.~D.~Polosa and V.~Riquer, Phys. Rev. D \textbf{94} (2016), 054026; J.~Wu, Y.~R.~Liu, K.~Chen, X.~Liu and S.~L.~Zhu, Phys. Rev. D \textbf{94} (2016), 094031; R.~Zhu, Phys. Rev. D {\bf 94} (2016) 054009.

\bibitem{Chen:2017dpy}
W.~Chen, H.~X.~Chen, X.~Liu, T.~G.~Steele and S.~L.~Zhu, Phys. Rev. D \textbf{96} (2017), 114017; Z.~G.~Wang, Eur. Phys. J. C \textbf{78} (2018), 518.

 \bibitem{Heller:1985cb} 
  L.~Heller and J.~A.~Tjon, Phys.\ Rev.\ D {\bf 32} (1985), 755; A.~V.~Berezhnoy, A.~V.~Luchinsky and A.~A.~Novoselov, Phys.\ Rev.\ D {\bf 86} (2012), 034004.
 
%\cite{Wu:2016vtq}
\bibitem{Wu:2016vtq} 
  J.~Wu, Y.~R.~Liu, K.~Chen, X.~Liu and S.~L.~Zhu, Phys.\ Rev.\ D {\bf 97} (2018), 094015;
  W.~Chen, H.~X.~Chen, X.~Liu, T.~G.~Steele and S.~L.~Zhu, Phys.\ Lett.\ B {\bf 773} (2017), 247; Y.~Bai, S.~Lu and J.~Osborne,
  arXiv:1612.00012 [hep-ph]; Z.~G.~Wang, Eur.\ Phys.\ J.\ C {\bf 77} (2017), 432;
  ~M.~N.~Anwar, J.~Ferretti, F.~K.~Guo, E.~Santopinto and B.~S.~Zou,
    Eur.\ Phys.\ J.\ C {\bf 78} (2018), 647; J.~M.~Richard, A.~Valcarce and J.~Vijande, Phys.\ Rev.\ D {\bf 95} (2017), 054019;
  A.~Esposito and A.~D.~Polosa, Eur.\ Phys.\ J.\ C {\bf 78} (2018), 782; M.~Karliner, S.~Nussinov and J.~L.~Rosner, Phys.\ Rev.\ D {\bf 95} (2017), 034011.
 
  \bibitem{Bedolla:2019zwg} 
  M.~A.~Bedolla, J.~Ferretti, C.~D.~Roberts and E.~Santopinto, arXiv:1911.00960 [hep-ph].  
 

\bibitem{GellMann:1964nj}
  M.~Gell-Mann,  Phys.\ Lett.\  {\bf 8} (1964) 214.
  
  \bibitem{Zweig:1964}
  G. Zweig, An SU3 model for strong interaction symmetry and its breaking, CERN TH-401, 1964.

\bibitem{Jaffe:1976ig}
  R.~L.~Jaffe,  Phys.\ Rev.\ D {\bf 15} (1977) 267; ibidem D \textbf{15} (1977), 281.

\bibitem{Jaffe:2004ph}
R.~L.~Jaffe,
Phys. Rept. \textbf{409} (2005), 1; F.~Wilczek, [arXiv:hep-ph/0409168 [hep-ph]].

\bibitem{Maiani:2004vq}
L.~Maiani, F.~Piccinini, A.~D.~Polosa and V.~Riquer,
Phys. Rev. D \textbf{71} (2005), 014028.

 \bibitem{Weinstein:1990gu}
J.~D.~Weinstein and N.~Isgur, Phys. Rev. D \textbf{41} (1990), 2236; F.~E.~Close and N.~A.~Tornqvist, J. Phys. G \textbf{28} (2002), R249-R267. 
 
 \bibitem{DeRujula:1976zlg}
A.~De Rujula, H.~Georgi and S.~L.~Glashow,
Phys. Rev. Lett. \textbf{38} (1977), 317.
 
 
 \bibitem{Choi:2007wga}
S.~Choi \textit{et al.} [Belle],
Phys. Rev. Lett. \textbf{100} (2008), 142001.

\bibitem{Aaij:2014jqa}
R.~Aaij \textit{et al.} [LHCb],
Phys. Rev. Lett. \textbf{112} (2014), 222002.

\bibitem{Ablikim:2013wzq}
M.~Ablikim \textit{et al.} [BESIII],
Phys. Rev. Lett.  \textbf{110} (2013) 252001; Phys. Rev. Lett. \textbf{111} (2013), 242001.
 

\bibitem{Bondar}
A. Bondar {\it et al.} [Belle Collaboration],
Phys.\ Rev.\ Lett. {\bf 108}, 122001 (2012).
 

 
 %\cite{Ali:2019roi}
\bibitem{Ali:2019roi}
  A.~Ali, L.~Maiani and A.~D.~Polosa,
  {\it Multiquark Hadrons}, Cambridge University Press (2019); A.~Esposito, A.~Pilloni and A.~D.~Polosa, Phys.\ Rept.\  {\bf 668} (2017) 1.
 
  \bibitem{Aaij:2015tga}
R.~Aaij \textit{et al.} [LHCb],
%``Observation of $J/\psi p$ Resonances Consistent with Pentaquark States in $\Lambda_b^0 \to J/\psi K^- p$ Decays,''
Phys. Rev. Lett. \textbf{115} (2015), 072001.

 \bibitem{Aaij:2019vzc}
R.~Aaij \textit{et al.} [LHCb], Phys. Rev. Lett. \textbf{122} (2019) 222001.

 \bibitem{Guo:2017jvc} 
  F.~K.~Guo, C.~Hanhart, U.~G.~Meissner, Q.~Wang, Q.~Zhao and B.~S.~Zou, Rev. Mod. Phys. \textbf{90} (2018) 015004.
 
  \bibitem{Li:2013ssa}
X.~Li and M.~B.~Voloshin, Mod. Phys. Lett. A \textbf{29} (2014), 1450060; E.~Braaten, C.~Langmack and D.~H.~Smith, Phys. Rev. D \textbf{90} (2014), 014044.

\bibitem{Swanson:2006st}
E.~S.~Swanson, Phys. Rept. \textbf{429} (2006), 243.

\bibitem{Becchi:2020uvq}
C.~Becchi, A.~Giachino, L.~Maiani and E.~Santopinto, [arXiv:2006.14388 [hep-ph]].

\bibitem{Godfrey:1985xj}
S.~Godfrey and N.~Isgur, Phys. Rev. D \textbf{32} (1985), 189.

			
\end{thebibliography}
\end{document}